\documentclass{pasj00}

\usepackage{times} 
\newcommand{\FigWidth}{66mm}

\begin{document}

\SetRunningHead{M. Choi}{B335 Millimeter Imaging}
\Received{2007/03/26}
\Accepted{2007/07/15}

\title{Observations of B335 in the Millimeter Continuum
       and the 226 GHz H$_2$CO Line}
\author{Minho Choi}
\affil{International Center for Astrophysics,
       Korea Astronomy and Space Science Institute,
       Hwaam 61-1, Yuseong, Daejeon 305-348, South Korea}
\email{minho@kasi.re.kr}


\maketitle

\begin{abstract}
The protostar B335 was observed in the 1.3 mm continuum
and in the H$_2$CO 3$_{12}$ $\rightarrow$ 2$_{11}$ line
with an angular resolution of about 8$''$.
The mass of the inner envelope detected by the dust continuum emission
is about 0.02~$M_\odot$.
The H$_2$CO spectrum at the protostellar position
shows a blue-skewed double peak profile,
suggesting that the kinematics of the inner envelope
is dominated by infall motion.
When the blueshifted and the redshifted peaks were imaged separately,
however, there is a small east-west displacement
between the maximum positions.
This displacement suggests that some part of the H$_2$CO emission
might come from the outflowing gas.
A combined effect of the infalling envelope and the outflow
on the radiative transfer is discussed.
This effect can make the line profile asymmetry
severer than what is expected from infall-only models.
\end{abstract}

\section{Introduction}

B335 is an isolated globule at a distance of 250 pc from the Sun
(Barnard 1927\footnote{See http://www.library.gatech.edu/barnard/
and the Dark Objects Catalogue therein.};
Bok et al. 1971; Bok \& McCarthy 1974;
Tomita et al. 1979; Frerking et al. 1987).
In the dense core of B335,
there is a low-luminosity (3 $L_\odot$) young stellar object
detected in far-IR/submillimeter
(Keene et al. 1983; Gee et al. 1985; Chandler et al. 1990).
A compact source elongated roughly in the north-south direction
was revealed by interferometric imaging
of the millimeter continuum emission from dust
(Hirano et al. 1992; Chandler \& Sargent 1993;
Choi et al. 1999; Wilner et al. 2000).
B335 is an archetypical Class~0 source
and considered as a protostar in the main accretion phase
(Andr{\'e} et al. 1993).
It is one of the best studied protostars.

Early radio observations of B335 suggested
that the 22~$M_\odot$ globule contains dense molecular gas
(Minn \& Greenberg 1973; Milman et al. 1975; Martin \& Barrett 1978).
Zhou et al. (1990) showed
that the dense core has a steep density gradient.
Star forming activities were revealed by molecular line observations
(Frerking \& Langer 1982; Frerking et al. 1987).
CO line observations showed
that there is a bipolar outflow oriented nearly in the plane of the sky
with an opening angle of $\sim$50$^\circ$
(Hirano et al. 1988; Cabrit et al. 1988).
The outflow axis is in the east-west direction,
and the millimeter continuum source is located
between the two outflow lobes,
which suggests that the protostar at the center of the dense core
is driving the outflow
(Hirano et al. 1992; Chandler \& Sargent 1993; Choi et al. 1999).
A compact radio thermal jet was detected near the protostellar position
(Anglada et al. 1992; Reipurth et al. 2002).
Herbig-Haro objects (HH 119) and shock-excited molecular hydrogen features
were detected along the outflow axis (Reipurth et al. 1992; Hodapp 1998).

B335 is one of the first examples of protostars
showing a spectroscopic signature of infall motion,
i.e., blue-skewed double-peak line profiles.
By comparing observations of several transitions of CS and H$_2$CO
with synthetic spectra
from radiative transfer calculations of self-similar collapse models,
parameters of the collapsing envelope,
such as infall radius and molecular abundances,
were derived (Zhou et al. 1993; Choi et al. 1995).
More detailed observations and modelings
using high-density tracers and near-IR extinction maps
provided a rich array of information on the protostellar collapse of B335
(Gregersen et al. 1997; Saito et al. 1999; Choi et al. 1999;
Wilner et al. 2000; Harvey et al. 2001; Takakuwa et al. 2007).
Interferometric observations of the millimeter continuum showed
that the density gradient of the inner envelope
has a power-law index of $\sim$1.55,
which is consistent with the theoretical expectations (1.5)
of gravitational free fall (Harvey et al. 2003a, 2003b).
Evans et al. (2005) presented chemical models
by comparing observations of about 30 transitions
with model spectra generated from calculations of dust radiative transfer,
molecular line radiative transfer, and chemical evolutions.
These observations and modelings collectively showed
that B335 is probably the best object
for the study of protostellar collapse and related phenomena,
and it is important to collect more data
in order to test our understanding of star formation
at the earliest evolutionary stage.

This Letter presents observations of the B335 region
in the $\lambda$ = 1.3 mm continuum and the H$_2$CO line
using the Berkeley-Illinois-Maryland Association (BIMA) array.
In \S~2 we describe our observations.
In \S~3 we report the results of the BIMA imaging
and discuss the structure of the protostellar core.

\section{Observations}

The $\lambda$ = 1.3 mm continuum
and the H$_2$CO 3$_{12}$ $\rightarrow$ 2$_{11}$ line (225.697772 GHz)
were observed using the BIMA array
in the D-array configuration on 1999 September 4.
The phase tracking center was
$\alpha_{2000}$ = 19$^{\mathrm h}$37$^{\mathrm m}$00\fs99
and $\delta_{2000}$ = 07\arcdeg34$'$10\farcs8.
A narrow spectral window was used for the H$_2$CO line
with a spectral resolution of 0.098 MHz
giving a velocity resolution of 0.13 km s$^{-1}$.
For the continuum,
six wide spectral windows at each sideband were used,
resulting in a single-sideband bandwidth of 600 MHz.
These wide windows do not contain any strong line.
The phase was determined by observing
a nearby quasar 2013+370 (NVSS J201528+371059).
The flux calibration was done by observing Uranus.
Maps were made using a CLEAN algorithm.
Natural weighting was used for all the images presented in this Letter.

\section{Results and Discussion}

\subsection{Millimeter Continuum}

The 1.3 mm continuum emission
was detected in both the lower and the upper sidebands.
Since combining the data of the two sidebands
did not improve the signal-to-noise ratio,
we present the map made using the lower sideband data only (Fig.~1).
The peak flux density is 97 $\pm$ 19 mJy beam$^{-1}$,
and the total flux is 180 $\pm$ 60 mJy.
(The peak flux of the upper-sideband map is 70 $\pm$ 20 mJy beam$^{-1}$.)
The 1.3 mm peak position is consistent
with the millimeter/centimeter position
of previously published maps (Wilner et al. 2000; Reipurth et al. 2002).

\begin{figure}[!b]
\begin{center}\FigureFile(\FigWidth,0){cont.eps}\end{center}
\caption{
Map of the $\lambda$ = 1.3 mm continuum toward the B335 region.
The contour levels are 40 and 80 mJy beam$^{-1}$.
Dashed contours are for negative levels.
The rms noise is 19 mJy beam$^{-1}$.
Shown at the bottom left corner is the synthesized beam:
FWHM = 10\farcs6 $\times$ 6\farcs7 and P.A. = 31\arcdeg.
The straight line at the bottom
corresponds to 2000 AU at a distance of 250 pc.
{\it Plus}:
Position of the 1.2 mm continuum source (Wilner et al. 2000).
{\it Filled circle}:
Position of the 3.6 cm continuum source (Reipurth et al. 2002).}
\end{figure}

 \begin{figure}[!t]
\begin{center}\FigureFile(\FigWidth,0){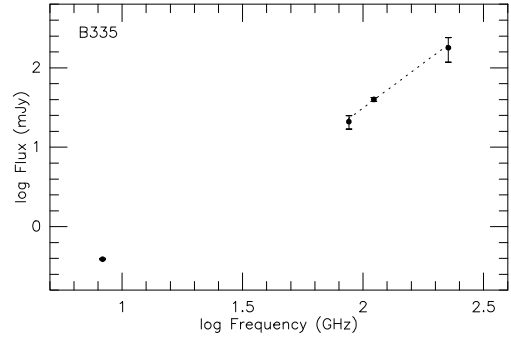}\end{center}
\caption{
Spectral energy distribution of B335 from interferometric observations.
Flux densities are from Reipurth et al. (2002),
Choi et al. (1999), Chandler \& Sargent (1993), and this work.
The flux density at $\lambda$ = 3.4 mm
from the data presented by Choi et al. (1999)
is 21 $\pm$ 4 mJy.
{\it Dotted straight line}:
Best-fit power-law spectrum in the millimeter range.
The spectral index is $\alpha$ = 2.3 $\pm$ 0.3.}
\end{figure}

Figure 2 shows the spectral energy distribution of B335
in the millimeter-centimeter range.
From the three data points in the millimeter band,
the best-fit power-law spectrum
has a spectral index of $\alpha$ $\approx$ 2.3,
which suggests that the millimeter emission mostly comes from dust.
For comparison, the spectral index in the submillimeter band
within a 40$''$ aperture is 2.9 $\pm$ 0.6 (Shirley et al. 2000).

To derive the mass from the dust continuum flux,
the dust emissivity given by Beckwith \& Sargent (1991) is assumed,
\begin{equation}
   \kappa_\nu = 0.1 \left({\nu\over{\nu_0}}\right)^\beta
                {\rm cm^2\ g^{-1}},
\end{equation}
where $\nu$ is the frequency, $\nu_0$ = 1200 GHz,
and $\beta$ is the opacity index
($\beta \approx \alpha - 2$, in the millimeter range).
If the emission is optically thin,
the mass can be estimated by
\begin{equation}
   M = {{F_\nu D^2}\over{\kappa_\nu B_\nu(T_d)}},
\end{equation}
where $F_\nu$ is the flux density, $D$ is the distance to the source,
$B_\nu$ is the Planck function, and $T_d$ is the dust temperature.
Dust radiative transfer modelings show
that the dust temperature within 0.006 pc (5$''$) from the center
ranges from 20 K to $\sim$40 K (Evans et al. 2005).
Assuming $D$ = 250 pc and $T_d$ = 30 $\pm$ 10 K,
a rough estimate of the mass from the interferometric observations
is $M$ = 0.021$^{+0.017}_{-0.007}$ $M_\odot$,
which includes the inner protostellar envelope and the disk.
The mass estimate is very sensitive to the opacity index,
and Chandler \& Sargent (1993) derived 0.21 $M_\odot$
by assuming $\beta$ = 1.2.
For comparison,
the 1.3 mm flux density within a 40$''$ beam is 0.57 Jy,
and the mass estimate within a 120$''$ beam is 1.2 $M_\odot$
(Shirley et al. 2000).

\subsection{The H$_2$CO Line}

Figure 3 shows the map of the H$_2$CO line
integrated over the full width of the line.
The strongest emission peak
coincides with the central protostellar position.
There is a secondary emission structure
near the northwestern corner of the map,
which is probably related with the outflow.

\begin{figure}[t]
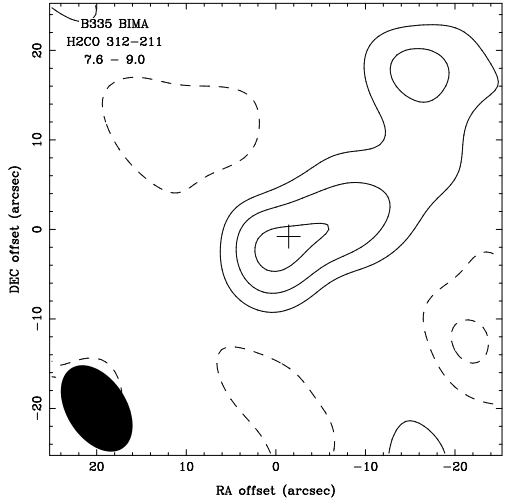

\begin{center}\FigureFile(\FigWidth,0){h2co.w.eps}\end{center}
\caption{
Map of the H$_2$CO line toward the B335 region.
The H$_2$CO line intensity was
averaged over the velocity interval
of $V_{\rm LSR}$ = (7.6, 9.0) km s$^{-1}$.
The contour levels are 1, 2, and 3 times 0.7 Jy beam$^{-1}$.
The rms noise is 0.26 Jy beam$^{-1}$.
Shown at the bottom left corner is the synthesized beam:
FWHM = 10\farcs6 $\times$ 6\farcs7 and P.A. = 32\arcdeg.
{\it Plus}:
Position of the 1.2 mm continuum source (Wilner et al. 2000).}
\end{figure}

Figure 4 shows the H$_2$CO spectrum at the protostellar position.
Similarly to most of the optically thick molecular lines seen toward B335,
the H$_2$CO line also shows a blue-skewed double peak line profile
that may be interpreted as a sign of infall motion.
However, when the blueshifted and the redshifted peaks
were imaged separately,
there is a small east-west displacement
between the maximum positions (Fig. 5).
The spectra at the maximum positions are shown in Figure 6.
Since the bipolar outflow is also in the east-west direction
(Hirano et al. 1992),
the displacement may be related with the outflow.
That is, some part of the H$_2$CO emission
might come from the outflowing gas.
This contribution from the outflow may provide an explanation
on the particularly poor fit
of the H$_2$CO 3$_{12}$ $\rightarrow$ 2$_{11}$ line
in the modeling of line profiles (Zhou et al. 1993; Choi et al. 1995).

\begin{figure}[t]
\begin{center}\FigureFile(\FigWidth,0){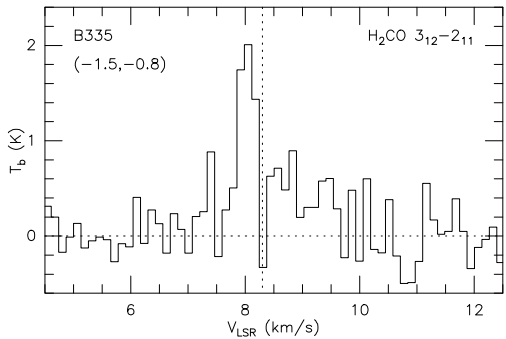}\end{center}
\caption{
Spectrum of the H$_2$CO line at the B335 protostellar position.
{\it Vertical dotted line}:
Systemic velocity of the B335 cloud core,
based on the average of optically thin lines
($V_{\rm LSR}$ = 8.30 $\pm$ 0.05 km s$^{-1}$; Evans et al. 2005).}
\end{figure}

 \begin{figure}[t]
\begin{center}\FigureFile(\FigWidth,0){h2co.br.eps}\end{center}
\caption{
Maps of the blueshifted (dashed contours)
and the redshifted (solid contours) peaks of the H$_2$CO line.
The H$_2$CO line intensity was
averaged over the velocity intervals
of $V_{\rm LSR}$ = (7.7, 8.2) and (8.4, 8.9) km s$^{-1}$, respectively.
The contour levels are 1, 2, and 3 times 1.3 Jy beam$^{-1}$.
The rms noise is 0.44 Jy beam$^{-1}$.}
\end{figure}

 \begin{figure}[!t]
\begin{center}\FigureFile(\FigWidth,0){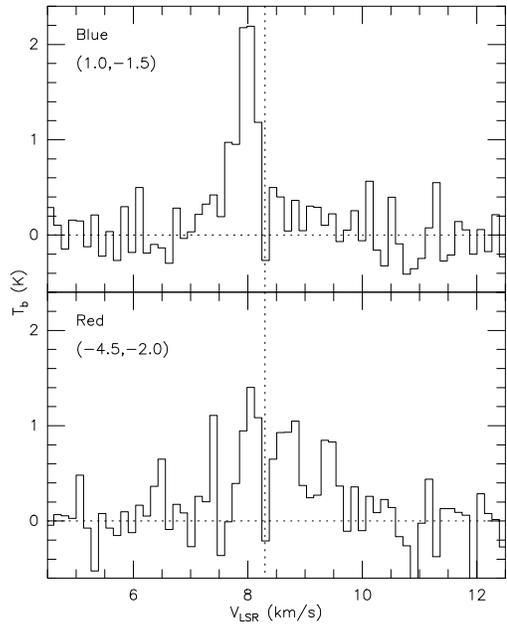}\end{center}
\caption{
Spectra of the H$_2$CO line at the maximum positions
of the blueshifted ({\it top}) and the redshifted ({\it bottom}) emission.
The position offset in arcseconds relative to the map center
is written at the top left corner of each panel (see Fig. 5).}
\end{figure}

Wilner et al. (2000) arrived at a similar conclusion
from interferometric observations of the CS $J$ = 5 $\rightarrow$ 4 line.
The high-resolution CS maps are dominated by clumps
aligned in the east-west direction,
which was interpreted as a slow, dense component of the bipolar outflow.
However, there is an interesting difference
between the H$_2$CO and the CS maps.
The strong blueshifted peak of the H$_2$CO emission
is located $\sim$3$''$ east of the protostar (Fig. 5),
while the strong blueshifted CS emission peak
is located $\sim$5$''$ southwest (Wilner et al. 2000).
If the blue asymmetry of the line profile
was caused entirely by a bright clump in the outflowing gas,
the strongest blueshifted emission
should have come from the same part of the outflow.
The difference between the H$_2$CO and the CS maps suggests
that the outflow alone is not necessarily the dominating cause
of the line profile asymmetry.
Since most of the optically thick molecular lines of B335
show a blue-skewed line profile (Evans et al. 2005),
the line asymmetry observed with single-dish telescopes
is probably caused by a systematic effect, most likely the collapse,
rather than the clumps in the outflow.
Especially, the blue-skewed profiles of off-center spectra
at positions in the north or south of the protostar
(i.e., at positions in the direction perpendicular to the outflow)
provide a strong evidence for infall motion
(Zhou 1995; Choi et al. 1999).

\begin{figure}[!b]
\begin{center}\FigureFile(\FigWidth,0){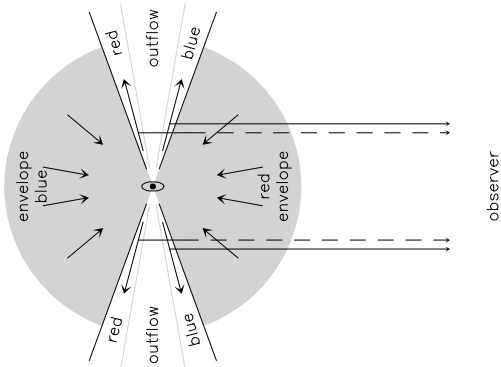}\end{center}
\caption{
Schematic diagram
showing the effect of the infalling envelope ({\it gray sphere})
on the emission from the outflow.
In this model, the outflow axis is nearly perpendicular to the line of sight,
the outflowing gas is optically thin,
and the infalling envelope is optically thick.
Radiation from the redshifted outflow ({\it dashed line})
can be absorbed by the front part of the infalling envelope.}
\end{figure}

While the difference between the CS and the H$_2$CO maps
may be caused by the chemical differentiation,
the H$_2$CO and the CS maps have two features in common.
First, the overall shape of the emission structure
is elongated in the east-west direction, parallel to the outflow axis.
Second, the blueshifted emission is
either stronger than or comparable to the redshifted one
in both the eastern and the western sides of the protostar.
One possible explanation of these features
may be the combined effect of the infalling envelope and the outflow
on the radiative transfer (Fig. 7).
If the outflow axis is nearly in the plane of the sky,
and if the outflowing gas and the infalling gas along a single line of sight
have comparable (projected) velocities,
the emission from the redshifted outflowing gas
can be absorbed by the redshifted infalling gas
in the front hemisphere of the envelope.
In contrast, the emission from the blueshifted envelope
may not suffer from absorption
if the blueshifted outflowing gas
is optically thin and/or more highly excited.
This effect can make the line profile asymmetry
severer than what is expected from infall-only models.
This model may not apply to every species/transitions
because the infalling and the outflowing gas
have quite different physical and chemical conditions.
For example, the enhancement of gas-phase H$_2$CO and CS
is expected in shocked outflowing gas
while some molecules (such as DCO$^+$)
are abundant only in the quiescent gas
(for example, Bachiller \& P{\'e}rez Guti{\'e}rrez 1997; Blake et al. 1995).
Such an outflow-envelope coupling in radiative transfer
may work for high-density tracer lines
that are optically thin in the outflowing gas
and moderately optically thick in the infalling envelope.
For a better understanding and comparisons with observations,
this model needs to be tested in detail
using multi-dimensional radiative transfer codes.

\bigskip

We thank KASI ISMJC members for helpful discussions.
This work was partially supported by the LRG Program of KASI.

%\vfill\centerline{\small\tt\input{.lt.stamp}}\vspace*{-\baselineskip}


\begin{thebibliography}{}
\bibitem[A(93)]{A93} Andr{\'e}, P., Ward-Thompson, D., \& Barsony, M. 1993,
                     ApJ, 406, 122
\bibitem[A(92)]{A92} Anglada, G., Rodriguez, L. F., Canto, J., Estalella, R.,
                     \& Torrelles, J. M. 1992, ApJ, 395, 494
\bibitem[B(97)]{B97} Bachiller, R., \& P{\'e}rez Guti{\'e}rrez, M. 1997,
                     APJ, 487, L93
\bibitem[B(27)]{B27} Barnard, E. E., Frost, E. B., \& Calvert, M. R. 1927,
                     A Photographic Atlas of Selected Regions in the Milky Way
                     (Washington: Carnegie Institute of Washington)
\bibitem[B(91)]{B91} Beckwith, S. V. W., \& Sargent, A. I. 1991,
                     ApJ, 381, 250
\bibitem[B(95)]{B95} Blake, G. A., Sandell, G., van Dishoeck, E. F.,
                     Groesbeck, T. D., Mundy, L. G., \& Aspin, C. 1995,
                     ApJ, 441, 689
\bibitem[B(71)]{B71} Bok, B. J., Cordwell, C. S., \& Cromwell, R. H. 1971,
                     in Dark Nebulae, Globules and Protostars, ed B. T. Lynds
                     (Tucson: Univ. Arizona Press), 33
\bibitem[B(74)]{B74} Bok, B. J., \& McCarthy, C. C. 1974, AJ, 79, 42
\bibitem[C(88)]{C88} Cabrit, S., Goldsmith, P. F., \& Snell, R. L. 1988,
                     ApJ, 334, 196
\bibitem[C(90)]{C90} Chandler, C. J., Gear, W. K., Sandell, G., Hayashi, S.,
                     Duncan, W. D., Griffin, M. J., \& Hazella, S. 1990,
                     MNRAS, 243, 330
\bibitem[C(93)]{C93} Chandler, C. J., \& Sargent, A. I. 1993,
                     ApJ, 414, L29
\bibitem[C(95)]{C95} Choi, M., Evans, N. J., II, Gregersen, E. M.,
                     \& Wang, Y. 1995, ApJ, 448, 742
\bibitem[C(99)]{C99} Choi, M., Panis, J.-F., \& Evans, N. J., II 1999,
                     ApJS, 122, 519
\bibitem[E(05)]{E05} Evans, N. J., II, Lee, J.-E., Rawlings, J. M. C.,
                     \& Choi, M. 2005, ApJ, 626, 919
\bibitem[F(82)]{F82} Frerking, M. A., \& Langer, W. D. 1982, ApJ, 256, 523
\bibitem[F(87)]{F87} Frerking, M. A., Langer, W. D., \& Wilson, R. W. 1987,
                     ApJ, 313, 320
\bibitem[G(85)]{G85} Gee, G., Griffin, M. J., Cunningham, T., Emerson, J. P.,
                     Ade, P. A. R., \& Caroff, L. J. 1985, MNRAS, 215, 15P
\bibitem[G(97)]{G97} Gregersen, E. M., Evans, N. J., II, Zhou, S.,
                     \& Choi, M. 1997, ApJ, 484, 256
\bibitem[H(01)]{H01} Harvey, D. W. A., Wilner, D. J., Lada, C. J.,
                     Myers, P. C., Alves, J. F., \& Chen, H. 2001,
                     ApJ, 563, 903
\bibitem[H(03)]{H03a} Harvey, D. W. A., Wilner, D. J., Myers, P. C.,
                      Tafalla, M., \& Mardones, D. 2003a, ApJ, 583, 809
\bibitem[H(03)]{H03b} Harvey, D. W. A., Wilner, D. J., Myers, P. C.,
                      \& Tafalla, M. 2003b, ApJ, 596, 383
\bibitem[H(92)]{H92} Hirano, N., Kameya, O., Kasuga, T., \& Umemoto, T. 1992,
                     ApJ, 390, L85
\bibitem[H(88)]{H88} Hirano, N., Kameya, O., Nakayama, M.,
                     \& Takakubo, K. 1988, ApJ, 327, L69
\bibitem[H(98)]{H98} Hodapp, K.-W. 1998, ApJ, 500, L183
\bibitem[K(83)]{K83} Keene, J., Davidson, J. A., Harper, D. A.,
                     Hildebrand, R. H., Jaffe, D. T., Loewenstein, R. F.,
                     Low, F. J., \& Pernic, R. 1983, ApJ, 274, L43
\bibitem[M(78)]{M78} Martin, R. N., \& Barrett, A. H. 1978, ApJS, 36, 1
\bibitem[M(75)]{M75} Milman, A. S., Knapp, G. R., Knapp, S. L.,
                     \& Wilson, W. J. 1975, AJ, 80, 101
\bibitem[M(73)]{M73} Minn, Y. K., \& Greenberg, J. M. 1973, A\&A, 22, 13
\bibitem[R(92)]{R92} Reipurth, B., Heathcote, S., \& Vrba, F. 1992,
                     A\&A, 256, 225
\bibitem[R(02)]{R02} Reipurth, B., Rodr{\'\i}guez, L. F., Anglada, G.,
                     \& Bally, J. 2002, AJ, 124, 1045
\bibitem[S(99)]{S99} Saito, M., Sunada, K., Kawabe, R., Kitamura, Y.,
                     \& Hirano, N. 1999, ApJ, 518, 334
\bibitem[S(00)]{S00} Shirley, Y. L., Evans, N. J., II, Rawlings, J. M. C.,
                     \& Gregersen, E. M. 2000, ApJS, 131, 249
\bibitem[T(07)]{T07} Takakuwa, S., Kamazaki, T., Saito, M., Yamaguchi, N.,
                     \& Kohno, K. 2007, PASJ, 59, 1
\bibitem[T(79)]{T79} Tomita, Y., Saito, T., \& Ohtani, H. 1979, PASJ, 31, 407
\bibitem[W(00)]{W00} Wilner, D. J., Myers, P. C., Mardones, D.,
                     \& Tafalla, M. 2000, ApJ, 544, L69
\bibitem[Z(95)]{Z95} Zhou, S. 1995, ApJ, 442, 685
\bibitem[Z(90)]{Z90} Zhou, S., Evans, N. J., II, Butner, H. M., Kutner, M. L.,
                     Leung, C. M., \& Mundy, L. G. 1990, ApJ, 363, 168
\bibitem[Z(93)]{Z93} Zhou, S., Evans, N. J., II, K{\"o}mpe, C.,
                     \& Walmsley, C. M. 1993, ApJ, 404, 232
\end{thebibliography}
\end{document}